# What Matters to Individual Investors:
# Price Setting in Online Auctions of P2P Consumer Loans

**Reto Rey[a]*, Andreas Dietrich[a]**

Version: October 22, 2022

**ABSTRACT:**

We analyze how retail investors price the credit risk of online P2P consumer loans in a reverse auction framework where personal interaction is absent. The explained interest rate variance is considerably larger than in comparable studies using bank loan data. This is unexpected, given the difference in experience and the limited set of information provided to the investor. Factors representing economic status significantly influence lender evaluations of the borrower's credit risk. Previous studies on gender discrimination found mixed results. We show that this is likely due to simplified specifications and find surprising results: male borrowers are charged a higher interest rates, conditional on being married and having children. Overall, our results indicate that retail investors exhibit a strong degree of predictability in this weakly regulated market for online P2P consumer loans.

Funding: This research did not receive any specific grant from funding agencies in the public, commercial, or not-for-profit sectors.

[a] Institute of Financial Services IFZ, Lucerne University of Applied Sciences, 6343 Rotkreuz, Switzerland
[*] Corresponding author / E-mail: reto.rey@hslu.ch / T: +41 41 757 67 13 / ORCID: 0000-0002-6591-6349

# 1 Introduction

A wide range of empirical studies in finance has shown the limitations of the traditional rationality paradigm (Barber et al., 2008; Barber and Odean, 2001, 2000; Barberis and Thaler, 2003; Kahneman, 2003; Kumar and Lee, 2006; Simon, 1956). This paper focuses on the pricing of unsecured P2P consumer loans and analyzes how retail investors assess risk, on the sole basis of information provided to them on an online platform by borrowers. The loan rates are set by the investors using a reverse auction mechanism, similar to the issuance of bonds. The market under observation is small, immature and dominated by retail investors with no specific qualification to price consumer loans. In addition to that, the asymmetry of information is potentially larger than in traditional lending schemes, as verified information on the borrower is scarce and personal interaction impossible.

But can we expect retail investors to analyze the data the environment provides in a sufficient way so they can make a reasonable decision? It has been suggested that since time and cognitive resources are limited, natural selection has designed minds that implement rules-of-thumb selectively to a subset of cues (Hirshleifer, 2001). But adequate pricing of loans is inevitable for the functioning of this credit market and affects the welfare of the borrower as well as investor. Among others, (Campbell, 2016) has emphasized the difficulty of the trade-off faced by financial regulators between the benefits of regulation and the cost of such regulation to other market participants. Therefore, it seems crucial to understand how the retail investors act in this so far weakly regulated market.[2]

In a first step we examine the extent to which differences in loan rates can be explained by the information available to the retail investor. In a second step, we examine the roots of the remaining dispersion. The dispersion of the observed loan rates - after taking into account available information - will serve as an indicator for the rationality of investors. In particular, we consider three categories of variables: (1) loan-specific, (2) borrower-specific and (3) macroeconomic factors. We were surprised to see our goodness of fit roughly twice as high as similar empirical studies using data on bank loan rates (Brick and Palia, 2007; Cerqueiro et al., 2011; Degryse and Ongena, 2005; Petersen and Rajan, 1994).

We were able to obtain the (internal and confidential) data on loan rates in the Swiss P2P market from 2008 to 2014 from *Cashare*, the most relevant Swiss P2P platform in this period. Our analysis of the lending rates differs in several important ways from previous studies. First, our data set from the Swiss market is unique as the relevant loan rates are not publicly available and to the best of our knowledge, no previous study has analyzed the Swiss P2P lending market. Even though still at a moderate level, the volume in the Swiss P2P lending market has increased exponentially in the past years. While the market volume was only CHF 100,000 in 2008, loans for CHF 186.7 million have been granted in 2017 (Dietrich and Amrein, 2018). Second, we investigate the role of additional and the interaction of borrower-specific information which have so far, to the best of our knowledge, been neglected in the P2P consumer lending market. In particular, there have been several empirical studies observing the effect of gender on the P2P lending rate (Berger and Gleisner, 2009; Chen et al., 2018; Herzenstein et al., 2011; Ravina, 2019), finding no clear evidence for significant differences between male and female borrowers. We augment this view by analyzing the interaction of gender, the marital status and the presence of children in the same household. Third and most important, interest rates from the observed P2P lending platform are – unlike for most other platforms – determined through an auctioning process. After controlling for the information provided by the borrower, this allows us to analyze the dispersion of the interest rates, which we see as an indicator for the rationality of investors with respect to their subjective expected risk-return tradeoff under imperfect information.

[2] One relevant regulation in the Swiss consumer credit market is the maximum interest rate of 10% (15% before 2016), according to the Federal Law on Consumer Credit.

The remainder of this paper is structured as follows: Section 2 provides a survey of the relevant academic literature; Section 3 contains a presentation of the basic model and the determinants of the P2P interest rates; Section 4 describes the data and methodology used to test our hypotheses; Section 5 presents the results from our empirical analysis; and Section 6 provides a summary and conclusions.

## 2 Related Literature

Online P2P lending, also referred to as crowdlending, loan based crowdfunding, or social lending can be separated into consumer and business lending. The development of the market, such as an increased involvement of institutional investors, especially in the US, has led to a broader use of the term marketplace lending. P2P financing activities currently exhibit high growth numbers. P2P consumer lending volumes in Europe (excluding the UK) grew tenfold from € 62.5 million issued loans to € 697 million by 2016 (Ziegler et al., 2018). The UK market was and remains around twice that size, growing from £ 127 million in 2012 to £ 1.4 billion in 2017 (Zhang et al., 2018). However, the issuance levels are still low compared to the £ 321.3 billion outstanding consumer loans to individuals at the end of 2017 (Bank of England, 2018). In this paper, we focus on the Swiss P2P consumer lending market which is still comparably small and therefore less interesting for institutional investors.

Since the inception of the first P2P lending platform Zopa in 2005 (Bachmann et al., 2011), an increasing amount of academic contributions have been published (Moritz and Block, 2016) provide a literature review) even though online P2P lending is a relatively young field of research. On the other hand, in contrast to the US or UK markets, there has been no empirical research yet which can be related to the Swiss P2P lending market.

The availability of comprehensive data from online P2P consumer lending platforms like *Prosper* and *Lending Club* has enabled a wide range of empirical research. Of major interest are factors affecting the funding success, lending rate and default rates. Specific focus has been laid on herding behavior (Herzenstein et al., 2011; Lee and Lee, 2012; Zhang and Liu, 2012), home bias (Lin and Viswanathan, 2015), trust (Duarte et al., 2012; Greiner and Wang, 2010), gender (Chen et al., 2018; Herzenstein et al., 2011; Pope and Sydnor, 2011; Ravina, 2019) and social networks (Freedman and Jin, 2014). (Agrawal et al., 2014) provide an economic interpretation of how transaction costs, reputation, and market design can explain the growth of crowdfunding. Table 1 gives an overview on relevant empirical findings.

The findings on gender are shown in the first five lines. Data from the US and China hints at smaller or similar interest rates for women than men. (Alesina et al., 2013) look at bank loan contracts to micro firms in Italy and find that women pay around 9 basis points more than men. On one hand, they attribute these differences to different negotiating skills. In our setting, negotiations skills are not relevant as we analyze an open auction setting with no bargaining. Their second explanation hints at taste-based discrimination.

**Table 1: Empirical findings of related literature on estimated changes in interest rates, funding probability and default rate**

| Article | Data Origin | Attribute | Interest Rate (%-pt.) | Funding Prob. (%-pt.) | Default Rate |
|---|---|---|---|---|---|
| Alesina et al., 2013 | Italy | Female (Bank Loans) | 0.09% | | |
| Bellucci et al., 2010 | Italy | Female (Bank loans to | = | - | |
| Barasinska & Schäfer, 2015 | Germany | Female | | = | |
| Pope & Sydnor, 2011 | USA | Female | -0.40% | 1.1% | |
| Herzenstein et al., 2011 | USA | Female | = | (-) | |
| Ravina, 2019 | USA | Female | = | 0.46% | = |
| Chen et al., 2018 | China | Female | -0.26% | = | - |
| Berger & Gleisner, 2009 | USA | Amount (per $ 1'000) | 0.25% | | |
| | | Grade (per Notch, AA - E) | 2.80% | | |
| | | Homeownership | = | | |
| | | Auction (vs. determined | -2.91% | | |
| | | Debt-to income Ratio | + | | |
| Chen et al., 2018 | China | Married | -0.17% | + | = |
| | | Mortgage loan | -0.27% | + | - |
| | | Punctuation marks (+10) | -0.08% | - | = |
| | | Words (+10) | 0.07% | + | = |
| Herzenstein et al., 2011 | USA | Married (vs. Single) | (+) | (-) | |
| | | Identity claim | | | |
| | | Economic Hardship | = | = | |
| | | Moral | = | = | |
| | | Religious | = | = | |
| | | Successful | -27%* | 23%* | |
| | | Trustworthy | -24%* | 33%* | |
| | | Hardworking | = | = | |
| Lin & Viswanathan, | USA | Amount | + | | + |
| 2016 | | Debt-to-Income | + | | + |
| | | Open Auction | - | | - |
| Pope & Sydnor, 2011 | USA | Asian | = | = | |
| | | Black | 0.80% | -26% | |
| | | Debt-to-Income Ratio | n.a. | -15% | n.a. |
| | | Hispanic | = | = | |
| | | Very attractive | = | = | |
| | | Very overweight | = | = | |

Table reports the estimated changes to the interest rate, funding probability and default rate for the dummy variables (if not else stated) in column 3. Numbers indicate percentage point changes, except for Herzenstein et al., where percent changes are reported (They did not use the absolute interest rate, but the deviation from the maximum rate a borrower is willing to pay). Equal sign stands for no or insignificant significant change, whereas for signs in brackets, significance is not verifiable.

An emerging issue is the interest rate setting mechanism of online P2P loans. There are currently two main regimes prevailing to determine the interest rate a borrower from a P2P lending platform has to pay: the reverse auction process and the posted prices. The reverse auction system, that works similar to the bond auctions where supply and demand determine interest rate, was widespread during the initial years of P2P lending. A potential borrower had to post his/her loan application on the platform and investors bid their investment amount with a corresponding minimal interest rate during the auction period. The Swiss P2P lending platform *Cashare*, the platform that has provided us with data for this paper, has been applying this auction procedure since its launch in 2008. Major players in the biggest P2P lending markets in the US and UK meanwhile apply the posted price process. In these models, the platform sets the interest rate for each loan listing based on the information available on the borrower. This simplifies the process for borrowers and lenders. *Prosper*, the first P2P lending platform in the US changed from the reverse auction regime to the posted-price regime in 2010, not long after their competitor *Lending Club* surpassed them in market share. A theoretical analysis concludes that reverse auction process, in contrast to posted prices (where the platform sets the interest rate for each borrower), fails to provide the cheapest outcome to the borrower (Chen et al., 2014). However, this theoretical model was contradicted by empirical evidence from the regime change of the P2P platform *Prosper* in December 2010 (Wei and Lin, 2015). They argue that the "wisdom of the crowd", as practiced in open auctions, may "allocate resources in a more socially desirable fashion". This is in line with results from auction theory. (Milgrom and Weber, 1982) in their seminal paper noted that competitive auctions can lead to higher bids, as bidders acquire useful information from their competitors. (Kagel and Levin, 1986) add that a larger number of bidders leads to "more aggressive" bidding, which in our case transforms into lower interest rates. Another conclusion from (Milgrom and Weber, 1982) is that the seller (lender) is incentivized to reveal full information, as it leads to a higher expected price (lower interest rate).

The discussion about the interest rate settings mechanism in crowdfunding platforms is especially interesting from a risk-return perspective. Professional lenders in the consumer loan business have become increasingly sophisticated over the last two decades with many lenders relying on data-driven models to assess and price credit risk (Staten, 2015). Credit scoring models allowed the lender to tailor the price and terms of a loan to a borrower's likelihood of repayment. In posted prices pricing models, crowdfunding platforms also set the interest rate for each borrower based on the traditional risk-return perspective and by trying to tailor the price based on the risk of the borrowers. In a reverse auction process however, as analyzed in this paper, risk and return are evaluated by the investors and it is not clear if these, usually not professional investors, correctly involve the risk-return tradeoff in their financial decisions.

So far and to the best of our knowledge, there is no recent paper specifically analyzing how rational investors are in their evaluation of risk and return and which factors determine the interest rates for P2P lending in a reverse auction process model. This is in part because most of the big platforms which are publishing their data are using the posted-price regime and therefore set the interest rates themselves. Additionally, for many platforms, loan and borrower-specific data is not publicly available. This paper addresses this issue and analyzes the drivers of the loan rates for P2P consumer loans. We analyze the determinants of P2P loan interest rates based on a unique dataset of all 665 P2P consumer loans granted by the largest Swiss platform with detailed single transaction information. We divide interest rate drivers into loan-specific, borrower-specific, and macroeconomic characteristics and determinants. The empirical research provided in this paper should thus make a valuable contribution to the literature on the P2P consumer loan market and especially to the question how non-professional investors price credit risk.

## 3 Data and Methodology

We estimate a multiplicative heteroscedasticity model proposed by (Harvey, 1976). This allows us to not only measure the direction and extent of how our variables affect the interest rate but let us also identify the determinants of their dispersion. The mean (1) and the variance (2) equations are given by:

$$\text{Rate}_{it} = \alpha + \beta_1 \times \text{Loan}_i + \gamma_1 \times \text{Borrower}_i + \delta_1 \times \text{Macro}_t + \varepsilon_i \quad (1)$$

$$\log(\sigma_i^2) = \alpha + \beta_2 \times \text{Loan}_i + \gamma_2 \times \text{Borrower}_i + \delta_2 \times \text{Macro}_t + \epsilon_i \quad (2)$$

*Rate* is the interest rate for loan *i* auctioned at time *t*; *Loan* is a vector of loan-specific variables for loan *i* as listed in Table 2. *Borrower* is a vector of the borrower-specific variables of loan *i*. *Macro* is a vector of macroeconomic variables at the time *t*. ε is an i.i.d. error term; and α, β, γ, and δ are vectors of parameters to be estimated.

We carry out this analysis for our full sample, and then separately for our two subsamples of data for years leading up to 2012, and from year 2013 onwards. The subsamples are defined this way based on two reasons. First, the market for P2P consumer loans has matured over time which is reflected in the higher volume and number of loans auctioned. Second, macroeconomic factors such as interest rates and unemployment had stabilized. With this reduced volatility, we expect a shift in the investors focus for determining their expected gross return, i.e., putting more focus on loan and borrower-specific factors. In an additional step, we also control for year and month of the loan auction to check whether there exists a time trend that is not accounted for.

All our unique loan-specific and borrower-specific data stem from *Cashare*, the biggest player in the Swiss P2P lending market with a market share of nearly 98% between 2008 and 2014. Therefore, our sample can be considered as representative for the whole P2P lending market in Switzerland during this time period. The sample for this dataset consists of information on 665 loans to private individuals granted between April 2008 and December 2014. Data for the monthly unemployment rate and the Swiss government bond rate are taken from the Swiss National Bank on a monthly basis, while data for the SMI on a daily basis is taken from the SIX Swiss Exchange.

## 4 Determinants of the Interest Rate

This section describes both, the dependent and the independent variables that we selected for our analysis of the P2P loan rates. Table 2 provides a summary of the variables as described below. As to our independent variables, our study focuses on three broad components to explain the interest rates of P2P consumer loans: the (1) loan-specific information, (2) borrower-specific information and (3) macroeconomic view.

The (1) loan-specific view analyzes elements such as loan volume and the loan period by investigating the effects of these elements on the interest rate for P2P consumer loans (see, e.g. (Dietrich, 2012). The (2) borrower-specific factors focus on aspects that affect a borrower's credit rating. In general, P2P lending is a classical principal-agent setting and faces the fundamental economic problem of information asymmetry. Asymmetric information arises because borrowers are better informed of their ability and willingness to repay than lenders, which in turn could lead to market failure (Akerlof, 1970; Stiglitz and Weiss, 1981). (Leland and Pyle, 1977) and (Campbel and Kracaw, 1980) suggest that informational asymmetries are a primary reason to explain the existence of financial institutions. In P2P lending, it is not easy for an individual lender, usually not an expert in analyzing and dealing with risks, to distinguish borrowers with a high probability of default from solvent ones. Furthermore, in the P2P consumer loan market there are no screening and signaling devices such as collateral and personal guarantees to distinguish the ex-ante riskiness of the borrower (Serrano-Cinca et al., 2015). Individual lenders are thus at a disadvantage and P2P lending is a risky activity for them. This information asymmetry could lead to adverse selection. In order to mitigate adverse selection, borrowers need quality information to adjust the interest rate they ask from the borrower, according to his risk profile. The P2P lending sites thus provide potential lenders with detailed information about borrowers and their loan purpose.

On the other side, the increasing popularity of P2P lending might be explained by the existence of transactions costs. P2P lending might lower search costs to both lenders and borrowers as it economizes on search (Stigler, 1961). Since the collection of deposits is bypassed, the lender is not subject to bank capital requirements and not overseen by bank regulators so far. According to (Maudos and De Guevara, 2004) operating costs are one of the most important factors in explaining bank interest rates and margins. Or as (Demirgüç-Kunt and Huizinga, 1999) formulate it, "banks pass on their operating costs to their depositors and lenders". The lower intermediation costs in the P2P lending might be transferred to clients in the form of lower interest rates for borrowers and higher revenues for lenders, compared to conventional financial institutions. In contrast to bank interest rates, operating costs are thus not a relevant driver of the interest rates for P2P consumer loans. Therefore, in our model we focus on variables that affect a borrower's credit rating such as the debt-to-income ratio, and additionally also consider rather subjective factors such as nationality and gender that were found to play a role in the decision-making process of the lender (Pope and Sydnor, 2011).

The (3) macroeconomic view perceives interest rates being driven by monetary policy changes and by economic cycles. An increase in the risk-free rate increases the interest rate on newly agreed loans. In addition, changes in business cycles, as measured by GDP growth or by the unemployment rate, may affect lending rates as the creditworthiness of borrowers varies over the business cycles (Bernanke and Gertler, 1995). We take this into account by including the seasonally adjusted unemployment rate, which according to Okun's law is strongly related to GDP (Okun, 1963). Furthermore, the labor market is highly relevant, given that we analyze consumer credits, which are not covered by collateral and thus depend on earned income.

### 4.1 Dependent Variable

Our dependent variable is the average interest rate paid by each borrower. It is set using reverse-auction mechanism by investors as described above. From an academic perspective, the reverse-auction regime is far more appealing than the posted-price regime, as it allows us to study the loan pricing behavior of the "crowd". On account of the wide range of variables from the dataset, we will be able to better understand key factors that play a role in lender's investment decisions, and whether the development and drivers of the P2P lending market have changed over time.

### 4.2 Independent Variables

This section describes the independent variables in our model. As Table 2 shows, we categorize these variables as (1) loan-specific factors, (2) borrower-specific variables and (3) macroeconomic factors. Our loan-specific factors consist of three variables. First, we investigate the relationship between loan volume (*loanamount*) and interest rate. As a risk-minimizing mechanism, many lenders bid small amounts on individual loans instead of placing one large bid with one borrower. We would thus expect that the larger the amount requested by the borrower, the higher the number of lenders needed to fund the auction completely. However, by law in Switzerland, a maximum of 20 persons were allowed to bid for one loan.[3] As a result, the larger the amount requested by a borrower, the higher an average bid per borrower, and thus, higher the perceived risk of the retail investors to lose money. Therefore, we expect a positive relationship between loan volume and interest rate.

Along with the conventional assumption of a normal yield curve, the loan period (*duration*) of a loan is expected to be positively correlated to the interest rate. The third loan-specific variable implies simple supply and demand assumptions. *Nrauctions90d* is a variable that reflects the loan demand as measured by the number of loan-auctions conducted in the 90 days before the loan is granted. We expect that loans auctioned during times of higher loan demand pay a higher interest rate.

Furthermore, we test nine different borrower-specific variables. First, we analyze whether the debt-to-income ratio (*debtincome*), a proxy for the credit rating, has an influence on the interest rate. It is calculated as a borrower's monthly installment arising from the loan he seeks in relation to his gross monthly income and indicates the resilience of a borrower towards unexpected expenses. Since consumers seek unsecured loans through P2P platforms, we can expect that the consumer's current income is positively associated with his or her ability to pay back the loan (Edelberg, 2006). A higher ratio is expected to negatively affect the creditworthiness of the borrower.

Additionally, we include dummy variables to control for gender (*dfemale*), nationality (*dswiss*), home ownership (*homeowner*), marital status (*married*) and children (*childrenyounger16*). We expect homeownership to be an indicator of higher net-worth, and thus, to be inversely related to the external finance premium as stated in (Bernanke et al., 1999). Regarding the gender variable regarded in isolation, we have no expectations. As the empirical evidence so far was mixed, we interact the gender variable simultaneously with marital status and children. This interaction should allow for a more distinct analysis of the investor's perception of a borrower's economic status. Assuming a subjective prejudice from a lenders perspective, an unmarried borrower with children might have to pay higher interest rates as they are expected to have higher regular expenses. Similarly, we expect Swiss nationals to pay a lower interest rate, since the unemployment rate for Swiss nationals has been smaller by factor 2.6 than the one for residents of Switzerland who are not citizens ("foreigners") during the past 25 years.

[3] Art. 6, Bundesgesetz über die Banken und Sparkassen (BankG).

**Table 2: Definition of Variables and Expected Sign**

| **Variable** | | **Description** | **Expected Sign** |
|---|---|---|---|
| Dependent | interest | Average interest rate on loan (%) | |
| Loan-specific | duration | Term of the loan (in months) | + |
| | loanamountk | Loan amount (in thousand CHF) | - |
| | nrauctions90d | Nr of loans for auction in the last 90 days | + |
| | interestmax | Maximum Interest Rate borrower is willing to pay | + |
| Borrower-specific | debtincome | Recurring monthly debt by gross monthly income | + |
| | dswiss | Dummy: Swiss borrower | - |
| | homeowner | Dummy: Homeowner | - |
| | married | Dummy: Marital status | = |
| | dfemale | Dummy: female borrower | = |
| | childrenyounger16 | Dummy: Has at least one child under 16 | + |
| Macro | unemployment[a] | Swiss unemployment rate (s.a.) at time of auction (%) | + |
| | ct_unempl[b] | Cantonal unemployment rate | + |
| | govbond3y[c] | 3 year Swiss Government Bond yield (%) | + |
| | SMI[d] | 3 month SMI performance (%) | - |
| | VSMI[d] | Volatility Index on the SMI | + |

[a, b,] : monthly data matched to the date of the auction end with a lag of 1 month

[c, d]: daily data matched to the date of the auction end

Since our dataset is spread over seven years, we control for changes in the macroeconomic environment. Including the seasonally adjusted unemployment rate (unemployment) allows us to control for business cycle effects. According to a wide range of empirical results from bank lending, we assume that the average probability of a loan default is highly correlated with the overall economic situation, and thus the unemployment rate affects the creditworthiness of borrowers as represented in interest rates (Ghosh, 2015; Jiménez et al., 2013; Louzis et al., 2012; Salas and Saurina, 2002; Sinkey and Greenawalt, 1991). We therefore expect this variable to be positively correlated with the loan rates. The coefficient of the government bond yield (*govbond3y*) as a measure for the risk-free interest rate is also expected to show a positive sign. We chose the 3-year yield to match the maturity of the majority of loans in the dataset. The average duration of a P2P loan in Switzerland is 32 months (see Table 3).

Since expectations about the future matter for investment decisions, we included the three-month performance of the Swiss Market Index (*SMI*) as a gauge for the investors' sentiment. People commonly use movements in equity prices as a leading indicator (Baker and Wurgler, 2006; Brown and Cliff, 2005, 2004). We thus infer that rising stock prices lower the risk premium asked by the loan investors.

### 4.3 Descriptive Statistics

Descriptive statistics for our full sample are shown in Table 3. The average interest rate is 9.8%, whereas the median is 9.7%, indicating that the distribution is not significantly skewed. The minimum interest rate is 1.9% and the maximum is 15%, as it was capped by the Federal Act on Consumer Credit.[4] The weighted average interest rate for all outstanding personal loans of the leading Swiss provider of consumer loans, Cembra Money Bank, was 11.33% in 2014 (Cembra Money Bank, 2015). In our subsample for the years 2013 and 2014 the average loan rate was 8.45%, so the average borrower paid 288 basis points less when taking a consumer loan from the online platform, as opposed to a loan from the largest market player. The average loan amount in our sample is 12,170 CHF, with a considerable positive skew. Three quarters of all loans amount to 15,704 CHF or less. The average loan term is 32 months. About one in four borrowers were female. This is comparable to the German lending platform Smava (Barasinska and Schäfer, 2014), but significantly lower than the US platform Prosper, where nearly half of the loans went to women (Ravina, 2019).

As in any functioning market, the supply of goods or services is relevant for the price setting. In our case, supply is the number and amount of loans requested. On average, 31.2 additional auctions were active during the three-month period before an observed loan was issued, with a maximum of 64 auctions. The minimum of 0 indicates that there were no other loans funded in three months before a loan was auctioned. This was - by definition - the case for the first ever loan auctioned, but not for any other of the observed loans.

As to our borrower-specific variables: 25% of the borrowers are female, 71% have a Swiss passport, and 21% of the borrowers are homeowners. The average debt to income ratio of the borrowers was at 7.37. The share of female borrowers decreased from 28% to 20%, whereas Swiss nationals account for 8 percentage points less in the last two years of the sample. Additionally, there are significantly more homeowners seeking a loan in 2013-2014, rising from 16% to 27%. The unemployment rate was slightly lower in 2013-2014 (3.16%) than 2008-2012 (3.29%) while the monthly performance of the Swiss Market Index (SMI) was more positive in 2013-2014 (2.98%) than in the years before (1.09%).

[4] Art. 1, Verordnung zum Konsumkreditgesetz (VKKG).

**Table 3: Descriptive Statistics**

| | (1) | | (2) | | (3) | |
|---|---|---|---|---|---|---|
| | Full Sample | | 2008 - 2012 | | 2013 – 2014 | |
| | mean | sd | mean | sd | mean | sd |
| **interest** | 9.80 | 2.63 | 11.00 | 2.62 | 8.45 | 1.89 |
| **duration** | 32.14 | 7.56 | 30.95 | 8.18 | 33.47 | 6.55 |
| **loanamountk** | 12.17 | 12.00 | 8.80 | 8.76 | 15.96 | 13.88 |
| **nrauctions90d** | 31.20 | 15.82 | 22.89 | 12.58 | 40.54 | 13.74 |
| **debtincome** | 11.02 | 19.36 | 10.24 | 21.03 | 11.91 | 17.27 |
| **dswiss** | 0.71 | 0.45 | 0.75 | 0.43 | 0.67 | 0.47 |
| **homeowner** | 0.21 | 0.41 | 0.16 | 0.37 | 0.27 | 0.44 |
| **married** | 0.39 | 0.49 | 0.31 | 0.46 | 0.48 | 0.50 |
| **dfemale** | 0.25 | 0.43 | 0.28 | 0.45 | 0.20 | 0.40 |
| **childrenyounger16** | 0.23 | 0.42 | 0.19 | 0.39 | 0.28 | 0.45 |
| **unemployment** | 3.24 | 0.36 | 3.32 | 0.49 | 3.16 | 0.03 |
| **ct_unempl** | 2.87 | 0.86 | 3.01 | 0.95 | 2.71 | 0.71 |
| **govbond3y** | 0.33 | 0.54 | 0.65 | 0.57 | -0.04 | 0.06 |
| **SMI** | 1.98 | 5.85 | 1.09 | 7.21 | 2.98 | 3.52 |
| **vsmi** | 17.97 | 5.39 | 20.90 | 5.88 | 14.68 | 1.57 |
| **interestmax** | 11.22 | 2.47 | 11.78 | 2.50 | 10.60 | 2.28 |
| **N** | 665 | | 352 | | 313 | |

This table reports descriptive statistics for the full sample and the two subsample periods.

# 5 Results

## 5.1 Full Sample Regression Analysis

Table 4 reports the regression results. We present two columns of results. Column 1 is considering loan-specific and macroeconomic variables, whereas column 2 additionally includes borrower-specific variables. This enables us to evaluate the relative explanatory power of the borrower-specific variables.

Our baseline model explains 50% of the variation in the interest rate of Swiss P2P consumer loans with all coefficients being significant. Including the borrower-specific variables, the model explains 54% of the interest rate variation. The sign and scope of the coefficients remain roughly stable across the two specifications. Even our baseline model with only six variables explains a comparably large share of interest rate variation across the 665 consumer loans placed on the platform and priced using an auction mechanism. We interpret this large goodness of fit as a sign for the investors being systematic in the way they price the consumer loans they invest in. In the augmented model, increasing the loan amount by CHF 10,000 leads to an expected interest rate increase of 35 basis points, compensating for the reduced diversification effect, i.e. higher notional default risk for a single investor.[5] A similar rate increase (37 basis points) is expected for loans that are auctioned during periods when the demand for loans in the past three months has been twice the average of 31. The duration of the loan is estimated to affect the interest rate positively, with its results being consistent with the assumptions of a normal yield curve. Each additional month in the duration can be associated with a 4.1 basis point rise in the interest rate.

[5] The maximum number of investors is, at this time, legally bound by 20 for each loan. For a loan of CHF 20,000 the average investment for each lender is CHF 1,000.

**Table 4: Full Sample Regression Results**

| Dependent Variable: interest rate | (1) | (2) | (3) | (4) |
|---|---|---|---|---|
| Duration | 0.037*** | 0.036*** | 0.042*** | 0.047*** |
| | (0.010) | (0.010) | (0.008) | (0.008) |
| Loan amount | 0.049*** | 0.048*** | 0.034*** | 0.036*** |
| | (0.008) | (0.008) | (0.006) | (0.006) |
| nrauctions90d | 0.036*** | 0.037*** | 0.005 | 0.042*** |
| | (0.007) | (0.007) | (0.006) | (0.007) |
| debtincome | 0.001 | 0.000 | 0.001 | 0.002 |
| | (0.004) | (0.003) | (0.002) | (0.002) |
| 1.dswiss | -0.480*** | -0.484*** | -0.290*** | -0.342*** |
| | (0.141) | (0.139) | (0.112) | (0.111) |
| 1.homeowner | -0.738*** | -0.738*** | -0.144 | -0.188 |
| | (0.175) | (0.175) | (0.133) | (0.127) |
| 1.married | 0.113 | 0.457** | 0.073 | 0.105 |
| | (0.169) | (0.190) | (0.166) | (0.153) |
| 1.dfemale | 0.203 | 0.115 | 0.047 | 0.028 |
| | (0.183) | (0.223) | (0.163) | (0.159) |
| 1.childrenyounger16 | 0.223 | -0.147 | -0.142 | -0.220 |
| | (0.197) | (0.498) | (0.343) | (0.293) |
| unemployment | 3.405*** | 3.318*** | 2.089*** | 1.528*** |
| | (0.257) | (0.256) | (0.180) | (0.254) |
| ct_unempl | 0.131 | 0.163* | 0.165** | 0.083 |
| | (0.092) | (0.089) | (0.070) | (0.068) |
| govbond3y | 1.914*** | 1.955*** | 1.417*** | 1.381*** |
| | (0.163) | (0.159) | (0.124) | (0.361) |
| SMI | 0.029* | 0.033** | 0.019 | 0.015 |
| | (0.016) | (0.017) | (0.013) | (0.013) |
| vsmi | 0.161*** | 0.158*** | 0.091*** | 0.073*** |
| | (0.021) | (0.023) | (0.017) | (0.022) |
| interestmax | | | 0.538*** | 0.497*** |
| | | | (0.035) | (0.037) |
| Constant | -7.799*** | -7.540*** | -7.328*** | -5.677*** |
| | (0.856) | (0.845) | (0.613) | (1.406) |
| Interactions (female, married, children) | No | Yes | Yes | Yes |
| Year Fixed Effects | No | No | No | Yes |
| Observations | 665 | 665 | 665 | 665 |
| Adjusted R-squared | 0.560 | 0.571 | 0.734 | 0.764 |

The table reports the coefficients and robust standard errors (in parenthesis) for the full sample regressions: First row for only the baseline estimate, and the second row for the augmented model. Sample size and R-squared are reported in the last two rows. Stars indicate the p-values: *** p<0.01, ** p<0.05, * p<0.1

The six borrower-specific variables also affect the interest rate in the expected direction. A 10-percentage point higher debt-to-income ratio is expected to result in a 31 basis points higher interest rate, a fairly moderate compensation for the higher default risk associated with borrowers of lower financial strength. (Berger and Gleisner, 2009) find a considerably larger effect (229.48 basis points) by using data from the US based platform Prosper. Owning a home, on the other hand, didn't result in a consistent significant impact in their data, whereas our results indicate a significant interest rate reduction of 72.9 basis points. This is contrasted by (Ramcharan and Crowe, 2013) who estimate an additional risk premium of approximately 50 basis points in association with homeownership. However, their estimations are based on a period of declining house prices in the US market (2006 to 2008) while Swiss real estate prices in our analyzed period were on the rise. Additionally, loan to value ratios tend to be held more conservatively in Switzerland.

Swiss passport holders, are estimated to pay 54 basis points less than foreigners living in Switzerland. This might seem as discrimination of foreign borrowers, but since the unemployment rate for Swiss nationals has been smaller by factor 2.6 than the one for foreigners living in Switzerland during the past 25 years this can be regarded as a rational risk premium. The coefficient for female borrowers is positive, but not statistically significant. This is in line with the existing literature on gender (see Table 1). However, this relation needs to be analyzed in a broader setting. Our results then show that women need to pay a significantly higher interest rate if they are unmarried and have at least one child under the age of 16 (+230 basis points). Married women with no children pay on average 53 basis points less than men (Table 5).

**Table 5: Marginal effect for women compared to men**

| | | | dy/dx | Std. Err. | [95% Conf. | Interval] |
|---|---|---|---|---|---|---|
| Full Sample | unmarried | no child | 0.25 | 0.23 | -0.21 | 0.70 |
| | | **child** | **2.30** | 0.79 | 0.75 | 3.86 |
| | married | no child | -0.53 | 0.38 | -1.27 | 0.21 |
| | | **child** | **0.76** | 0.60 | -0.42 | 1.94 |
| 2008 - 2012 | unmarried | no child | 0.26 | 0.30 | -0.32 | 0.84 |
| | | **child** | 2.15 | 0.88 | 0.41 | 3.89 |
| | married | no child | -1.01 | 0.59 | -2.18 | 0.15 |
| | | **child** | 1.82 | 0.60 | 0.64 | 3.00 |
| 2013 - 2014 | unmarried | no child | 0.11 | 0.27 | -0.42 | 0.64 |
| | | **child** | 3.08 | 0.68 | 1.74 | 4.41 |
| | married | no child | 0.16 | 0.38 | -0.60 | 0.91 |
| | | **child** | -0.90 | 0.36 | -1.62 | -0.18 |

The table reports the marginal effects for women compared to men, interacted with the marital status as well as children for the full sample and our two subsample periods 2008-2012 and 2013-2014. Additionally, we report the standard errors and 95% confidence interval.

The macroeconomic variables have a significant impact on the interest rates in our full sample. A rise in the seasonally adjusted unemployment rate by one standard deviation (0.35) transforms into an expected rise in the loan rate of 152 basis points. Similarly, the impact of one standard deviation rise in the 3-year government bond yield (0.54) is 127 basis points. The logic behind the strong responses to unemployment and the risk-free interest rate can be associated with the financial accelerator mechanism as described in (Bernanke et al., 1999). A rise in the three-month return of the SMI by 1% is associated with a 4.6 basis point lower interest rate.

### 5.2 Robustness and Subample Regression Analysis

Robust results were obtained when also correcting for month and year of loan auctions as shown in Table 8. Signs and magnitude of the coefficients were largely unchanged. For another test of whether there is a latent time trend that we did not account for in the regression, the actual and fitted values of the interest rates were plotted against time. As shown in Figure 1 in the appendix, the two series exhibit no systematic deviations.

To investigate the impact of market maturity and the changed macroeconomic environment, we split the sample into two time periods: the period from 2008-2012; and years 2013-2014. The two subsamples exhibit fairly robust results. Signs do not change, whereas the statistical and economic significance shifts among our three defined categories of explanatory variables.

Table 6 presents the estimates for the two subsample periods. Columns 1 and 2 show the baseline and augmented estimates for the subsample from 2013-2014. Columns 3 and 4 show the estimates for the subsample 2008-2012.

**Table 6: Subsample Regression Results**

| Dependent variable: Interest Rate | (1) | (2) | (3) | (4) |
|---|---|---|---|---|
| | Subsample 2013 - 2014 | | Subsample 2008 - 2012 | |
| Model: | Baseline | Augmented | Baseline | Augmented |
| loanamountk | 0.048*** | 0.033*** | 0.031*** | 0.025** |
| | (0.010) | (0.011) | (0.011) | (0.012) |
| duration | 0.074*** | 0.076*** | 0.034** | 0.029** |
| | (0.012) | (0.012) | (0.013) | (0.015) |
| nrauctions90d | 0.057*** | 0.054*** | 0.031*** | 0.029*** |
| | (0.007) | (0.007) | (0.010) | (0.010) |
| debtincome | | 0.055*** | | 0.026 |
| | | (0.020) | | (0.017) |
| dswiss | | -0.521*** | | -0.612*** |
| | | (0.160) | | (0.214) |
| homeowner | | -0.252 | | -1.114*** |
| | | (0.197) | | (0.284) |
| 1.married | | 0.237 | | 0.769*** |
| | | (0.226) | | (0.256) |
| 1.dfemale | | 0.059 | | 0.259 |
| | | (0.269) | | (0.297) |
| 1.childrenyounger16 | | -1.352*** | | 0.609 |
| | | (0.300) | | (0.538) |
| unemployment | -1.324 | -2.890 | 3.589*** | 3.487*** |
| | (3.572) | (3.432) | (0.302) | (0.281) |
| govbond3y | 0.972 | 0.969 | 1.293*** | 1.404*** |
| | (1.500) | (1.414) | (0.188) | (0.172) |
| SMI | -0.005 | -0.019 | -0.033* | -0.025 |
| | (0.029) | (0.028) | (0.017) | (0.016) |
| Constant | 7.136 | 12.234 | -3.659*** | -2.983*** |
| | (11.258) | (10.837) | (1.025) | (1.045) |
| Interactions (female, married, children) | No | Yes | No | Yes |
| Observations | 313 | 313 | 352 | 352 |
| Adjusted R-squared | 0.428 | 0.499 | 0.482 | 0.540 |

The table reports the coefficients and robust standard errors (in parenthesis) for the subsample regressions: First row for the baseline estimate (Corresponds to Column 1 of Table 4), and the second row for the augmented model (Corresponds to Column 2 of Table 4). Stars indicate the p-values: *** p<0.01, ** p<0.05, * p<0.1

As can be seen in Table 3, the macroeconomic environment was more stable during the second period. Furthermore, the P2P consumer lending market in Switzerland matured, indicated by a higher amount of loans granted and by the average number of loans auctioned within the three-month period. The number of auctions in this three-month time spam almost doubled from 22.9 in the first subsample, to 40.5 in the years 2013 and 2014.

Across the two subsamples, we see a clear shift in the explanatory power from macroeconomic variables towards loan- and borrower specific variables. Comparing columns (2) and (4), the coefficients for the loan duration and debt to income ratio have more than doubled. This is an indication that investors put more emphasis on traditional risk-related factors and thus become more rational.

Homeownership, on the other hand, has not been associated with significantly lower interest rates in the second subsample, as opposed to the period of 2008 – 2012 (-111 basis points). Another significant difference between the two subsamples is observed with respect to the borrower's gender. As expected, investors do not generally discriminate between male and female borrower. However, women with at least one child under the age of 16 are considered as riskier borrowers. In the first subsample, both, married and unmarried women with children are paying a higher risk premium. On average, these women had to paid roughly 200 basis points more than men with children (Table 5). In the second subsample, this only holds for *unmarried* women with children. Married women with children were even seen as a better risk than their male peers (-90 basis points). These findings indicate that investors do discriminate, but not solely on the base of gender, but according to economic status.

### 5.3 Variance Equation

We now look at the causes for dispersion in the interest rates. Table 7 shows the full sample as well as the subsample results of our variance equation, using the augmented model. Following equation (2), positive coefficients indicate a larger unexplained deviation of the mean equation due to the corresponding variable. Negative coefficients indicate a smaller residual variance. Thus, the significantly negative coefficient for homeowners in the full sample estimate indicates that for those borrowers, the unexplained variance is smaller. Or put differently, given the information set provided, tenant's borrowing rates are less predictable than those of homeowners.

Most noticeable is the diminishing impact on the dispersion. In the first subsample, five variables showed a significant alteration of dispersion. In this less mature market, Swiss citizens and married borrowers experienced a lower variation in their interest rate. Or put differently, foreigners and unmarried borrowers experience more unexplained variations in their borrower rates. This anomaly disappeared in the subsample of 2013 – 2014, when the market grew more mature and macroeconomic fluctuations were smaller. Only two variables are significant, both relating to the interaction terms of being married with children. Married men with children are exposed to more dispersion, while for married women with children the borrowing rate seems better predictable using the information available to the investor.

**Table 7: Results from the Variance Equation**

| Dependent variable: Error terms of the q | (1)<br>Full Sample | (2)<br>Subsample<br>2008 – 2012 | (3)<br>Subsample<br>2013 - 2014 |
|---|---|---|---|
| loanamountk | 0.002 | 0.020 | 0.013 |
| | (0.011) | (0.022) | (0.010) |
| duration | -0.020 | -0.030* | 0.026 |
| | (0.012) | (0.016) | (0.024) |
| nrauctions90d | -0.010 | -0.003 | -0.003 |
| | (0.007) | (0.011) | (0.011) |
| debtincome | -0.019 | -0.055 | -0.018 |
| | (0.018) | (0.035) | (0.024) |
| dswiss | 0.395* | -0.419* | 0.449 |
| | (0.219) | (0.238) | (0.282) |
| homeowner | -0.554** | -0.353 | -0.241 |
| | (0.268) | (0.338) | (0.387) |
| 1.married | 0.259 | -0.950** | -0.226 |
| | (0.294) | (0.383) | (0.420) |
| 1.dfemale | 0.872*** | 0.357 | 0.141 |
| | (0.244) | (0.337) | (0.445) |
| 1.childrenyounger16 | 1.059** | -0.743 | -1.265 |
| | (0.457) | (0.668) | (0.772) |
| unemployment | 0.216 | 0.144 | 9.914 |
| | (0.263) | (0.317) | (6.291) |
| govbond3y | 0.427** | -0.285 | 0.485 |
| | (0.198) | (0.209) | (2.262) |
| SMI | 0.008 | -0.013 | 0.021 |
| | (0.021) | (0.019) | (0.043) |
| Constant | -0.548 | 1.245 | -33.302* |
| | (1.050) | (1.148) | (20.083) |
| Interactions (female, married, children) | Yes | Yes | Yes |
| Observations | 665 | 352 | 313 |
| Adjusted R-squared | 0.046 | 0.028 | 0.010 |

The table reports the coefficients and robust standard errors (in parenthesis) for the variance equation of the full sample as well as the subsample regressions using the augmented model. Sample size and R-squared are reported in the last two rows. Stars indicate the p-values: *** p<0.01, ** p<0.05, * p<0.1

# 6 Conclusion

This paper offers insights into the pricing mechanism of the P2P consumer loan market, where retail investors set prices in a reverse auction. We use a unique dataset of 665 P2P consumer loans with detailed single transactions and including specific information about the loan and borrowers in Switzerland from April 2008 to December 2014 and evaluate how consistent investment decisions are. By analyzing the determinants of the interest rates demanded by investors in this purely supply and demand driven loan market with imperfect information, we shed light on the question of how retail investors handle the risk and return trade-off in a competitive open auction environment. In particular, this dataset allows us to examine to what extent differences in loan rates are a function of (1) loan-specific, (2) borrower-specific and (3) macroeconomic factors.

The loan-specific variables produce statistically, as well as economically significant results. We find that interest rates for loans are higher if the duration is longer, if the loan amount is larger or if there are more loan auctions in the same period and as a result, more opportunities for retail investors to participate in this alternative market. The signs of our coefficients are as expected and imply that the interest rate setting mechanism seems to be rather rational. We also find that the macroeconomic environment significantly influences the interest rates for P2P consumer loans. Loan rates are higher when the general interest level and unemployment rate is high. These are further indicators that retail investors generally act rather rational and demand higher interest rates if the risk-free interest rate and the economic uncertainty is higher.

Furthermore, we find that borrower-specific factors, representing the main credit risk of the P2P consumer loans, also significantly affect the interest rates. Due to the anonymous nature of the internet based lending market, granting a loan is characterized by high risk and uncertainty. On P2P lending platforms, individuals try to differentiate themselves by providing signals of trustworthiness. Key variables, such as the borrower's economic status, significantly influence lender evaluations of the borrower's credit risk and thus the interest rates. Our results show that interest rates are significantly higher when the debt-to-income ratio is higher.

However, we also find some indication of discrimination by the lenders. A first indicator for discrimination is based on passport holdings. Swiss passport holders pay significantly lower interest rates than foreigners living in Switzerland although, e.g., the debt-to-income level for Swiss is, on average, even higher. Another discrimination effect comes from the gender and the presence of children living in the same household. On an aggregate level, we find no significant differences in interest rates between men and women, which is in line with the existing literature. The coefficient for female borrowers in our full sample as well as the two subsamples are positive, but not statistically significant. However, taking a deeper look shows that women need to pay a significantly higher interest rate if they are unmarried and have at least one child under the age of 16 (+230 basis points). On the other hand, married women with no children pay on average 53 basis points less than their men peers (Table 5). This difference disappears in the second subsample where the market matures.

But what persists across all samples is the higher interest rate asked from unmarried women with children. So not gender itself affects the retail investor's expected risk-return tradeoff, but rather some interpretation of economic status. So despite some discriminatory aspects, our analysis indicates that investors act rather rational according to their beliefs. We also see clear signs of a maturing P2P lending market, as investors shift their focus towards more established risk-related factors such as loan duration and debt to income ratio.

# 7 Appendix

**Table 8: Robustness – Regression Including Months and Years**

| | Full Sample | | 2008 – 2012 | | 2013 - 2014 | |
|---|---|---|---|---|---|---|
| | (1) | (2) | (3) | (4) | (5) | (6) |
| VARIABLES | coef | se | coef | se | coef | se |
| loanamountk | 0.034*** | (0.009) | 0.041*** | (0.013) | 0.031** | (0.013) |
| duration | 0.052*** | (0.010) | 0.034** | (0.014) | 0.076*** | (0.012) |
| nrauctions90d | 0.070*** | (0.010) | 0.047*** | (0.015) | 0.082*** | (0.021) |
| debtincome | 0.042*** | (0.014) | 0.027 | (0.017) | 0.055** | (0.024) |
| dswiss | -0.586*** | (0.136) | -0.563*** | (0.217) | -0.459*** | (0.170) |
| homeowner | -0.618*** | (0.175) | -1.035*** | (0.290) | -0.271 | (0.207) |
| 1.married | 0.491*** | (0.172) | 0.542** | (0.245) | 0.289 | (0.219) |
| 1.dfemale | 0.222 | (0.200) | 0.344 | (0.286) | 0.009 | (0.267) |
| 1.married#1.dfemale | -0.593 | (0.384) | -0.746 | (0.683) | 0.107 | (0.466) |
| 1.childrenyounger16 | -0.180 | (0.412) | 0.643 | (0.594) | -1.242*** | (0.280) |
| 1.married#1.childrenyounger16 | -0.007 | (0.460) | -1.051 | (0.676) | 1.182*** | (0.372) |
| 1.dfemale#1.childrenyounger16 | 2.166*** | (0.687) | 2.043** | (1.036) | 2.773*** | (0.746) |
| 1.married#1.dfemale#1.childrenyounger16 | -1.606* | (0.905) | -0.200 | (1.373) | -3.834*** | (0.965) |
| unemployment | 2.124*** | (0.430) | 2.436*** | (0.460) | -3.491 | (3.766) |
| govbond3y | -0.007 | (0.477) | -0.471 | (0.571) | 0.572 | (2.020) |
| SMI | -0.011 | (0.014) | -0.005 | (0.018) | -0.055 | (0.034) |
| 2009.year | -0.127 | (0.980) | -1.071 | (1.052) | | |
| 2010.year | -1.954* | (1.079) | -2.515** | (1.213) | | |
| 2011.year | -2.464** | (1.003) | -3.076*** | (1.171) | | |
| 2012.year | -3.438*** | (1.257) | -4.516*** | (1.490) | | |
| 2013.year | -5.436*** | (1.269) | | | | |
| 2014.year | -6.041*** | (1.410) | | | -0.950* | (0.558) |
| 2.month | 0.456 | (0.339) | 1.050** | (0.450) | -0.616 | (0.435) |
| 3.month | 0.506 | (0.312) | 1.581*** | (0.438) | -0.575 | (0.353) |
| 4.month | 0.053 | (0.342) | 1.080** | (0.479) | -1.295*** | (0.442) |
| 5.month | -0.094 | (0.327) | 0.356 | (0.486) | -0.436 | (0.463) |
| 6.month | -0.363 | (0.362) | -0.362 | (0.598) | -0.725 | (0.521) |
| 7.month | -0.096 | (0.361) | 0.347 | (0.511) | -0.935 | (0.627) |
| 8.month | -0.235 | (0.372) | 0.153 | (0.524) | -1.368* | (0.700) |
| 9.month | 0.264 | (0.358) | 0.202 | (0.555) | -0.142 | (0.566) |
| 10.month | -0.358 | (0.329) | -0.310 | (0.453) | -0.980* | (0.526) |
| 11.month | -0.316 | (0.325) | -0.385 | (0.466) | -0.369 | (0.407) |
| 12.month | -0.351 | (0.347) | -0.415 | (0.494) | -0.672 | (0.536) |
| Constant | 2.479 | (1.691) | 3.147 | (2.051) | 14.366 | (12.02) |
| Observations | 665 | | 352 | | 313 | |
| Adjusted R-squared | 0.637 | | 0.581 | | 0.517 | |

Robust standard errors in parentheses
*** p<0.01, ** p<0.05, * p<0.1

**Figure 1** **Actual Versus Fitted Interest Rates Over Time**

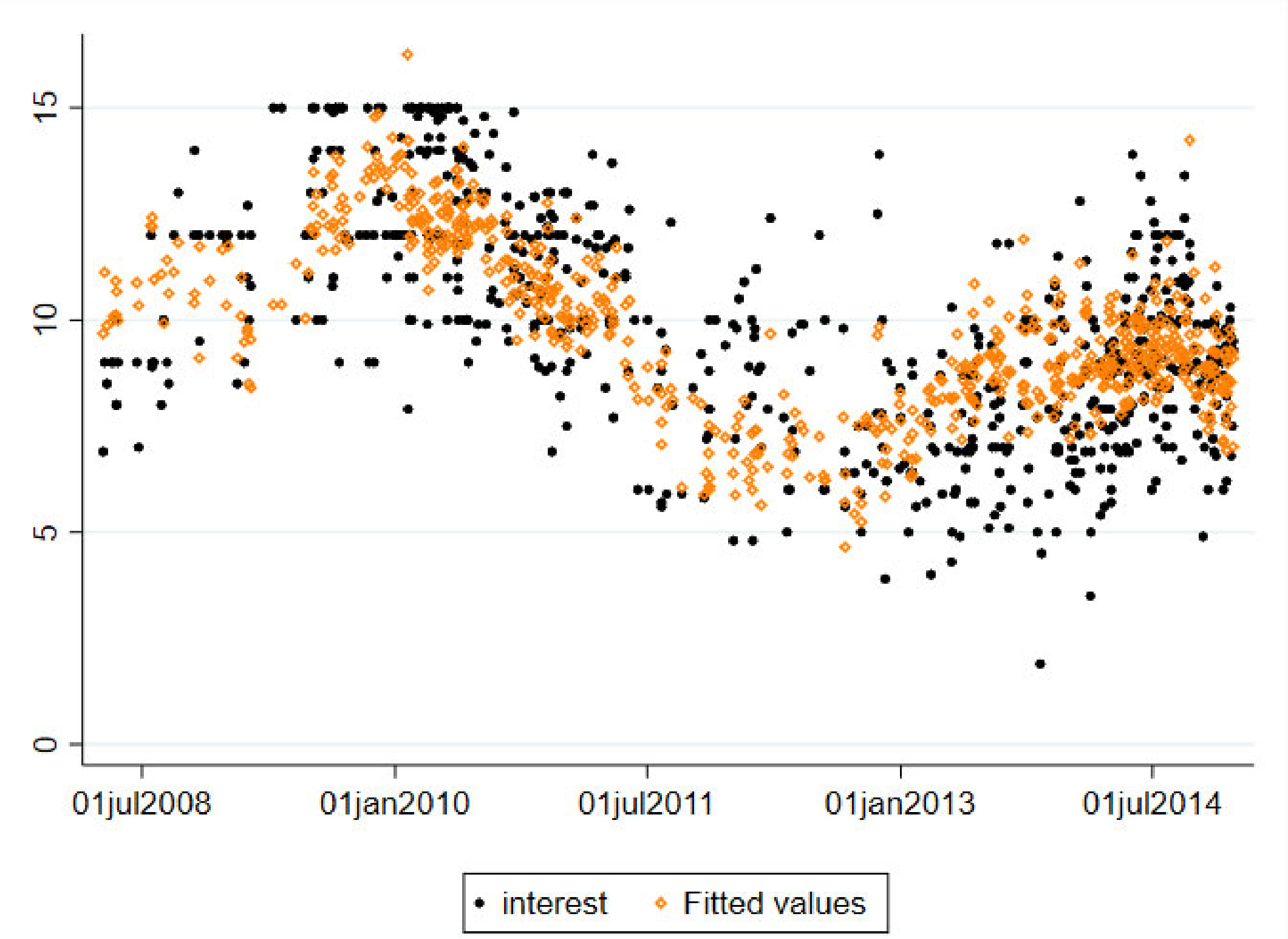